\documentclass[12pt]{article}
\usepackage{amsmath}

\newcommand{\bm}{\begin{multiline}}
\newcommand{\beq}{\begin{equation}}
\newcommand{\eeq}{\end{equation}}
\newcommand{\beqs}{\begin{eqnarray}}
\newcommand{\eeqs}{\end{eqnarray}}

\newcommand{\ra}{\rightarrow}

\begin{document}

\thispagestyle{empty}

\begin{flushright}
hep-th/yymmxxx
\end{flushright}

\hfill{}

\hfill{}

\hfill{}

\vspace{32pt}

\begin{center}

\textbf{\Large New Multiply Nutty Spacetimes }

\vspace{48pt}
{\bf Robert Mann}\footnote{E-mail: 
{\tt mann@avatar.uwaterloo.ca}}
{\bf and Cristian Stelea}\footnote{E-mail: 
{\tt cistelea@uwaterloo.ca}}

\vspace*{0.2cm}

{\it $^{1}$Perimeter Institute for Theoretical Physics}\\
{\it 31 Caroline St. N. Waterloo, Ontario N2L 2Y5 , Canada}\\[.5em]

{\it $^{1,2}$Department of Physics, University of Waterloo}\\
{\it 200 University Avenue West, Waterloo, Ontario N2L 3G1, Canada}\\[.5em]
\end{center}

\vspace{30pt}

\begin{abstract}
We construct new solutions of the vacuum Einstein field equations with
multiple NUT parameters, with and without cosmological constant. These
solutions describe spacetimes with non-trivial topology that are
asymptotically $dS$, $AdS$ or flat. We also find the the multiple nut
parameter extension of the inhomogeneous Einstein metrics on complex line
bundles found recently by L\"u, Page and Pope. We also provide a more general
form of the Eguchi-Hanson solitons found by Clarkson and Mann. We discuss
the global structure of such solutions and possible applications in string
theory.
\end{abstract}

\vspace{32pt}

\setcounter{footnote}{0}

\newpage

\section{Introduction}

Despite their unusual properties, Taub-NUT spacetimes have received
increased attention in recent years. Part of their attractiveness issues
from the reputation NUT-charged spacetimes have of being a `counterexample
to almost anything' in General Relativity \cite{Misner}. For example such
spacetimes typically (but not always) have closed time-like curves (CTCs) in
their Lorentzian section. This apparently less-than-desirable feature
actually makes them more interesting. They have served as testbeds for the
AdS/CFT conjecture \cite{Hawking, Mann1, Chamblin, Lorenzo}. Even more
recently a very interesting result was obtained via a non-trivial embedding
of the Taub-NUT geometry in heterotic string theory, with a full conformal
field theory definition (CFT) \cite{Johnson1}. It was found that nutty
effects were still present even in the exact geometry, computed by including
all the effects of the infinite tower of massive string states that
propagate in it. This might be a sign that string theory can very well live
even in the presence of nonzero NUT charge, and that the possibility of
having CTCs in the background can still be an acceptable physical situation.

Furthermore, in asymptotically dS settings, regions near past/future
infinity do not have CTCs, and NUT-charged asymptotically dS spacetimes have
been shown to yield counter-examples to some of the conjectures advanced in
the still elusive dS/CFT paradigm \cite{strominger}- such as the maximal
mass conjecture and Bousso's entropic N-bound conjecture \cite%
{rick1,rick2,rickreview}. Moreover, they have been used to uncover deep
results regarding gravitational entropy \cite{Hawking,Mann1,Hunter,Robinson}
and, in particular, they exhibit breakdowns of the usual relation between
entropy and area \cite{micky1} (even in the absence of Misner string
singularities). Furthermore, it has been recently noted that the boundary
metric Lorentzian sector of these spaces in AdS-backgrounds is in fact
similar with the G\"odel metric \cite{micky2,micky3}.

The first solution in four dimensions describing such an object was
presented in \cite{Taub,NUT}. Intuitively a NUT charge corresponds to a
magnetic type of mass. Its presence induces a so-called Misner singularity
in the metric, analogous to a `Dirac string' in electromagnetism \cite
{Misner}. This singularity is only a coordinate singularity and can be
removed by choosing appropriate coordinate patches. However, expunging this
singularity comes at a price: in general we must make coordinate
identifications in the spacetime that yield CTCs in certain regions.

There are known extensions of the Taub-NUT solutions to the case when a
cosmological constant is present and also in the presence of rotation \cite%
{Demiansky,Gibbons:nf,Ortin,Klemm}. In these cosmological settings, the
asymptotic structure is only locally de Sitter (for a positive cosmological
constant) or anti-de Sitter (for a negative cosmological constant) and we
speak about Taub-NUT-(a)dS solutions.

Generalizations to higher dimensions follow closely the four-dimensional
case \cite{Bais,Page,Akbar,Robinson,Awad,Lorenzo,csrm,Page1,Chen,Gauntlett}.
In constructing these metrics the idea is to regard Taub-NUT spacetimes as
radial extensions of $U(1)$ fibrations over a $2k$-dimensional base space
endowed with an Einstein-K\"ahler metric $g_{B}$. Then the $(2k+2)$%
-dimensional Taub-NUT spacetime has the metric: 
\begin{equation}
ds^{2}=F^{-1}(r)dr^{2}+(r^{2}+N^{2})g_{B}-F(r)(dt+NA)^{2}  \label{l1}
\end{equation}
where $t$ is the coordinate on the fibre $S^{1}$ and the one-form $A$ has
curvature $J=dA$, which is proportional to some covariantly constant $2$%
-form. Here $N$ is the NUT charge and $F(r)$ is a function of $r$.

Recently NUT-charged spacetimes with more than one nut parameter have been
obtained as exact solutions to the higher-dimensional Einstein equations %
\cite{csrm}. However explicit solutions were provided only up to seven-dimensions. These higher-dimensional spaces are constructed as radial extensions of circle fibrations over even-dimensional base spaces that can be factored in the form $B=M_{1}\times \dots \times M_{d-2}$, where 
$M_{i}$ are two dimensional spaces of constant curvature. One can then
associate a NUT charge $N_{i}$ for every such two-dimensional factor. The
above metric ansatz would be then modified by replacing $(r^{2}+N^{2})g_{B}$
with the sum $\sum_{i}(r^{2}+N_{i}^{2})g_{M_{i}}$, while $2NA$ would be
replaced by $\sum_{i}2N_{i}A_{i}$. For example we can use the sphere $S^{2}$%
, the torus $T^{2}$ or the hyperboloid $H^{2}$ as factor spaces. These
solutions represent the generalizations of the spacetimes studied in refs. 
\cite{Bais,Robinson,Awad,Lorenzo}.

Another class of solutions introduced in \cite{csrm} used a generalised
ansatz: 
\begin{equation}
ds^{2}=F^{-1}(r)dr^{2}+(r^{2}+N^{2})g_{M}+\alpha r^{2}g_{Y}-F(r)(dt+2NA)^{2}\label{g2}
\end{equation}
in which one constructs the higher dimensional Taub-NUT space as a
generalised fibration over an Einstein-K\"ahler manifold $M$. The
non-trivial feature of this ansatz is that now the fibre contains besides
the $(r,t)$-sector a general Einstein space $Y$, endowed with an Einstein
metric $g_{Y}$. This type of solutions was later generalised to arbitrary
dimensions by L\"u, Page and Pope in \cite{Page1}.

In this letter we generalise both of these types of solutions to
include multiple NUT parameters in arbitrary dimensions. We first describe
the generalisation of the ansatz (\ref{l1}) for an arbitrary number of
factors $M_{i}$ in the factored form of the base space $B$. To analyse the
possible singularities of these metrics we switch over to their Euclidian
sections by performing analytic continuations of the time coordinate $t$ and of
the nut parameters. Following \cite{Page} we go on to analyse the regularity
constraints to be imposed on these Euclidian sections in order to obtain
regular metrics that can be extended globally to cover the whole manifold.
We find that for generic values of the parameters these metrics are
singular: it is only for a sagacious choice of the parameters that they
become regular. As an example of this general analysis we focus on the
six-dimensional case and we explicitly consider the cases of a Taub-NUT-like
fibration over the base spaces $S^{2}\times S^{2}$ and $CP^{2}$.

In Section $3$ we present a more general form of the solution (\ref{l1}) in
which we replace the $2$-dimensional factors $M_{i}$ by arbitrary
even-dimensional Einstein-K\"ahler manifolds. We use here the
normalisation that the Ricci tensor for each manifold $M_{i}$ can be written
as $Ricci(M_{i})=\delta _{i}g(M_{i})$. For each factor $M_{i}$ we associate
a nut parameter $N_{i}$. Consistent with what was conjectured in \cite{csrm}, we find that generically there are constraints to be imposed on the possible values of the cosmological constant $\lambda $, the nut parameters $N_{i}$ and the values of the various $\delta $'s. These solutions represent
the multiple nut parameter generalisation of the inhomogeneous Einstein metrics on complex line-bundles described in \cite{Page}. We find that we can cast these solutions in another form that explicitly encodes the constraint conditions into the metric.

In Section $4$ we present the multiple nut parameter extension of the
metrics (\ref{g2}) constructed by L\"u, Page and Pope \cite{Page1}. In this
case we replace the Einstein-K\"ahler manifold $M$ by a product of
Einstein-K\"ahler manifolds $M_{i}$ with arbitrary even-dimensions
and to each such factor we associate a nut parameter $N_{i}$. The case in
which $Y$ is one-dimensional is particularly interesting to us since it will
provide us with the general form of the odd-dimensional
Eguchi-Hanson-type solitons found recently by Clarkson and Mann \cite{EH}.

The final section is dedicated to conclusions and possible applications.

Our conventions are: $(-,+,...,+)$ for the (Lorentzian) signature of the
metric; in even $d$ dimensions our metrics will be solutions of the vacuum
Einstein field equations with cosmological constant $\Lambda =\pm \frac{%
(d-1)(d-2)}{2l^{2}}$, which can be expressed in the form $G_{ij}+\Lambda
g_{ij}=0$ or in the equivalent form $R_{ij}=\lambda g_{ij}$, where $\lambda =%
\frac{2\Lambda }{d-2}=\pm \frac{d-1}{l^{2}}$. By an abuse of terminology we will still call $\lambda$ cosmological constant.

\section{The general solution}

We assume that the $(d-2)$-dimensional base space in our construction can be
factored as a product of $p$ factors, $B=M_{1}\times \dots \times M_{p}$
where $M_{i}$ are $2$-dimensional spaces of constant curvature normalised
such that Ricci$(M_{i})=\delta _{i}g(M_{i})$. The metric ansatz that we use
is then: 
\begin{equation}
ds_{d}^{2}=-F(r)(dt+\sum\limits_{i=1}^{p}2N_{i}A_{i})^{2}+F^{-1}(r)dr^{2}+%
\sum\limits_{i=1}^{p}(r^{2}+N_{i}^{2})g_{M_{i}}  \label{I2}
\end{equation}%
where 
\begin{equation*}
A_{i}=\Biggl\{%
\begin{array}{cl}
\cos \theta _{i}d\phi _{i}, & ~~~~\mathrm{{for~~\delta =1~(sphere)}\nonumber}
\\ 
\theta _{i}d\phi _{i}, & ~~~~\mathrm{{for~~\delta =0~(torus)}\nonumber} \\ 
\cosh \theta _{i}d\phi _{i}, & ~~~~\mathrm{for~~\delta =-1~(hyperboloid)}%
\end{array}%
\end{equation*}

By solving the vacuum Einstein equations with cosmological constant we
obtain: 
\begin{equation}
F(r)=\frac{r}{\prod\limits_{i=1}^{p}(r^{2}+N_{i}^{2})}\bigg[%
\int\limits^{r}\left( \delta _{1}-\frac{d-1}{l^{2}}(s^{2}+N_{1}^{2})\right) 
\frac{\prod\limits_{i=1}^{p}(s^{2}+N_{i}^{2})}{s^{2}}ds-2m\bigg]
\label{Frgeneral}
\end{equation}%
while the constraints on the values of the nut parameters $N_{i}$ and the
cosmological constant $\lambda $ can be expressed in the very simple form
for every $i,j=1\dots p$: 
\begin{equation}
\lambda (N_{j}^{2}-N_{i}^{2})=\delta _{j}-\delta _{i}
\end{equation}

If all the factors $M_{i}$ coincide then we can satisfy this constraint in
two ways: either we can take all the nut parameters to be equal $N_{i}=N$
and keep the cosmological constant non-vanishing or else we can take $\lambda =0$ and keep
the nut parameters independent. However, if at least two factors $M_{i}$ are
different, it is inconsistent to set $\lambda =0$. In this case all the nut
parameters corresponding to identical $M_{i}$ factors must remain equal,
while those corresponding to different $M_{i}$ factors must remain distinct
such that the above constraints are still satisfied.

Note that while the factors in front of the $M_{i}$ are never zero, this is
not so for the Euclidean section of the metric. Therefore, when we shall
address the possible singularities of the above metrics we shall focus
mainly on their Euclidean sections, recognising that the Lorentzian versions
are singularity-free -- apart from quasi-regular singularities \cite{Konk},
which correspond to the end-points of incomplete and inextensible geodesics
that spiral infinitely around a topologically closed spatial dimension.
However, since the Riemann tensor and all its derivatives remain finite in
all parallelly propagated orthonormal frames we take the point of view that
these represent some of mildest of types of singularities and we shall
ignore them when discussing the singularity structure of the Taub-NUT
solutions. We also note that for asymptotically $dS$ spacetimes that have no
bolts quasi-regular singularities are absent \cite{Anderson}.

Scalar curvature singularities have the possibility of manifesting
themselves only in the Euclidean sections. These are simply obtained by the
analytic continuations $t\rightarrow i\tau$ and $N_{j}\rightarrow in_{j}$,
and can be classified by the dimensionality of the fixed point sets of the
Killing vector $\xi =\partial /\partial \tau$\ that generates a $U(1)$
isometry group. In four dimensions, the Killing vector that corresponds to
the coordinate that parameterizes the fibre $S^{1}$ can have a
zero-dimensional fixed point set (we speak about a `NUT' solution in this
case) or a two-dimensional fixed point set (referred to as a `bolt'
solution). The classification in higher dimensions can be done in a similar
manner. If this fixed point set dimension is $\left( d-1\right) $ the
solution is called a Bolt solution; if the dimensionality is less than this
then the solution is called a NUT solution. If $d=3$, Bolts have dimension $%
2 $ and NUTs have dimension $0$. However if $d>3$ then NUTs with larger
dimensionality can exist \cite{csrm,Page1}. Note that fixed point sets need
not exist; indeed there are parameter ranges of NUT-charged asymptotically $%
dS$ spacetimes that have no Bolts \cite{Anderson}.

Since a singularity analysis for some of the lower-dimensional cases of
these metrics has previously appeared \cite{csrm} we shall limit ourselves
to an outline its general features. The analysis is a direct application of
the one given in \cite{Page}. In order to extend the local metrics presented
above to global metrics on non-singular manifolds the idea is to turn all
the singularities appearing in the metric into removable coordinate
singularities. For generic values of the parameters in the solution the
singularities are not removable, corresponding to conical singularities in
the manifold. We are mainly interested in the case of compact Einstein-K\"ahler manifolds $M_{i}$. Generically the K\"ahler forms $J_{i}$ on $M_{i}$ can be equal to $dA_{i}$ only locally. Hence we need to use a number of overlapping coordinate patches to cover the whole manifold. In
order to render the $1$-form $d\tau +\sum 2n_{i}A_{i}$ well-defined we need
to identify $\tau $ periodically. In general this can be done if the ratios
of all the parameters $n_{i}$ are rational numbers. If we choose them to be
positive integers we can define $q=\gcd \{n_{1},\dots ,n_{p}\}$ and require
the period of $\tau $ to be given by: 
\begin{equation}
\beta =\frac{8\pi q}{k}  \label{beta}
\end{equation}%
where $k$ is a positive integer. It is also necessary to eliminate the
singularities in the metric that appear as $r$ is varied over $M$. Attention
must be paid to the so-called endpoint values of $r$: these are the values
for which the metric components become zero or infinite. For a complete
manifold $r$ must range between two adjacent endpoints -- if any conical
singularities occur at these points they must be eliminated. The finite
endpoints occur at $r=\pm n_{i}$ or at the simple zeros of $F_{E}(r)$. In
general $r=\pm n_{i}$ are curvature singularities unless $F_{E}=0$ there as
well. To eliminate a conical singularity at a zero $r_{0}$ of $F_{E}(r)$ we
must restrict the periodicity of $\tau $ to be given by: 
\begin{equation}
\beta =\frac{4\pi }{|F_{E}^{\prime }(r_{0})|}
\end{equation}%
and this will generally impose a restriction on the values of the parameters
once we match it with ($\ref{beta}$). For compact manifolds the radial
coordinate takes values between two finite endpoints and the regularity
constraint must be imposed at both endpoints. If the manifold is noncompact
then the cosmological constant is non-positive and the radial coordinate
takes values between one finite endpoint $r_{0}$ and one infinite endpoint $%
r_{1}=\infty $. For our asymptotically locally flat or $(A)dS$ solutions the
infinite endpoints are not within a finite distance from any points $r\neq
r_{1}$ so there is no regularity condition to be imposed at $r_{1}$. In this
case the regularity conditions to be satisfied are that $F_{E}(r)>0$ for $%
r\geq r_{0}$ and $\beta =\frac{4\pi }{|F_{E}^{\prime }(r_{0})|}$.

Consider for example the six-dimensional case with a fibration over the base
space $S^{2}\times S^{2}$. If the cosmological constant is non-zero, $%
\lambda =-5/l^{2}$, then we must have $n_{1}=n_{2}=n$. Regularity of the $1$%
-form $d\tau -2nA$ forces the periodicity of $\tau $ to be given by $8\pi n/k
$, where $k$ is an integer. We must match this periodicity with the one
emerging by requiring absence of conical singularities at the root $r_{0}$
of $F_{E}(r)$, which is
\begin{equation}
F_{E}(r)=\frac{%
3r^{6}+(l^{2}-15n^{2})r^{4}-3n^{2}(2l^{2}-15n^{2})r^{2}-6mrl^{2}-3n^{4}(l^{2}-5n^{2})%
}{3l^{2}(r^{2}-n^{2})^{2}}  \label{F6dim}
\end{equation}%
from the Einstein equations using (\ref{Frgeneral}). The nut
solution corresponds to $r_{0}=n$ in which case we obtain $\frac{4\pi }{%
\left| F_{E}^{\prime }(r)\right| }=12\pi n$. As there is no integer $k$ for
which the periodicities can be matched, we conclude that this solution is
singular. Indeed it is easy to check that $r_{0}=n$\ is the location of a
curvature singularity. To define a bolt solution it is sufficient to require 
$r_{0}>n$ and the regularity condition in this case is given by $\frac{4\pi 
}{\left| F_{E}^{\prime }(r)\right| }=\frac{8\pi n}{k}$, with $k$ an integer.
Solving this constraint we find 
\begin{equation*}
r_{0}=\frac{kl^{2}\pm \sqrt{k^{2}l^{4}-80n^{2}l^{2}+400n^{4}}}{20n}.
\end{equation*}%
If the cosmological constant vanishes then we can have different values for
the nut parameters. Without loss of generality, assume that $n_{1}>n_{2}$
and that they are rationally related. Then it is easy to see that, in order
to keep the metric positive definite, we have to restrict the range of the
radial coordinate such that $r>n_{1}$. As above, the periodicity of the $%
\tau $ coordinate is found to be $8\pi n_{2}/k$, where $k$ is an integer. We
have to match this with the periodicity imposed on $\tau $ by eliminating
the conical singularities at a root $r_{0}$ of $F_{E}(r)$. We distinguish
two types of solutions: a nut and a bolt. The nut solution corresponds to $%
r_{0}=n_{1}$ and in this case the periodicity $\frac{4\pi }{|F_{E}^{\prime
}(n_{1})|}=8\pi n_{1}$ cannot be matched with $8\pi n_{2}/k$ for any integer
value of $k$. However note that $r_{0}=n_{1}$ is not the location of a
curvature singularity! On the other hand, the bolt solution corresponds to $%
r\geq r_{0}>n_{1}$ and the periodicity is found to be $\frac{4\pi }{%
|F_{E}^{\prime }(n_{1})|}=\frac{8\pi n_{1}}{p}$, where $p$ is some integer.
It is now possible to match it with $8\pi n_{2}/k$ with $k$ an integer such
that $p/k=n_{1}/n_{2}$. The bolt solution is then non-singular.

The situation changes considerably if we take $B=CP^{2}$ as the base space %
\cite{Bais,Page,Awad}. In this case $p=1$ in eq. (\ref{g2}) and the
submanifold $g_{M_{1}}$ has the metric 
\begin{equation}
d\Sigma _{2}^{~2}=\frac{du^{2}}{\left( 1+\frac{\delta u^{2}}{6}\right) ^{2}}+%
\frac{u^{2}}{4\left( 1+\frac{\delta u^{2}}{6}\right) ^{2}}\left( d\psi +\cos
(\theta )d\phi \right) ^{2}+\frac{u^{2}}{4\left( 1+\frac{\delta u^{2}}{6}%
\right) }(d\theta ^{2}+\sin ^{2}\theta d\phi ^{2})  \label{CP2Metric}
\end{equation}%
with $F_{E}(r)$ still\ given by (\ref{F6dim}). However the
one-form $2nA$ is now given by 
\begin{equation}
A=\frac{u^{2}n}{2\left( 1+\frac{\delta u^{2}}{6}\right) }\left( d\psi +\cos
\theta d\phi \right) 
\end{equation}%
We need to find the smallest value of $\int 2ndA$ over a closed 2-chain.
Changing coordinates so that $u=\sqrt{\frac{6}{\lambda }}\tan \chi $ the $%
CP^{2}$ metric can be written as \cite{Akbar2}%
\begin{eqnarray}
ds^{2} &=&\frac{6}{\delta }\left( d\chi ^{2}+\frac{\sin ^{2}\chi }{4}\left(
d\theta ^{2}+\sin ^{2}\theta d\phi ^{2}\right) +\sin ^{2}\chi \cos ^{2}\chi
\left( d\psi +\cos \theta d\phi \right) ^{2}\right)   \label{CP2alt} \\
A &=&\frac{3n}{2\delta }\sin ^{2}\chi \left( d\psi +\cos \theta d\phi
\right)   \notag
\end{eqnarray}%
where $0\leq \chi \leq \frac{\pi }{2}$, $0\leq \theta \leq \pi $, $0\leq
\phi \leq 2\pi $, and $0\leq \psi \leq 4\pi $. \ We see from (\ref{CP2alt})
that $\chi =0$ is a `nut' in this subspace, and so there is no closed
2-chain on which to integrate $2ndA$. However at $\chi =\frac{\pi }{2}$ the $%
\left( \theta ,\phi \right) $ sector is a 2-dimensional bolt. Hence at $\chi
=\frac{\pi }{2}$ we obtain 
\begin{equation*}
\int 2ndA=2\frac{3n}{2\delta }4\pi =\frac{12\pi n}{\delta }
\end{equation*}%
implying that the periodicity of $\tau $ can be $12\pi n/k$, where we use
the normalisation $\delta =1$. Equating this to $\frac{4\pi }{\left|
F_{E}^{\prime }(r=n)\right| }=12\pi n$ yields\footnote{%
The parameter $m=\frac{4n^{3}(6n^{2}-l^{2})}{3l^{2}}$ is fixed by requiring
that $F_{E}(n)=0$.} $k=1$, and the geometry at $r_{0}=n$ is smooth. Thus we
can obtain regular nut and bolt solutions if the base space is $CP^{2}$. More generally, for $CP^{q}$ the periodicity is $\frac{4\pi n\left(q+1\right) }{k\delta }$, with $k$ an integer \cite{Robinson,Page}.

\section{A more general class of solutions}

In this section we present a more general class of Taub-NUT metrics in even
dimension. These spaces are constructed as complex line bundles over a
product of Einstein-K\"ahler spaces $M_{i}$, with dimensions $2q_{i}$
and metrics $g_{M_{i}}$. Then the total dimension is $d=2(1+\sum%
\limits_{i}^{p}q_{i})$. The metric ansatz that we use is the following: 
\begin{equation*}
ds_{d}^{2}=-F(r)(dt+\sum\limits_{i=1}^{p}2N_{i}A_{i})^{2}+F^{-1}(r)dr^{2}+%
\sum\limits_{i=1}^{p}(r^{2}+N_{i}^{2})g_{M_{i}}
\end{equation*}%
Here $J_{i}=dA_{i}$ is the K\"ahler form for the $i$-th
Einstein-Kahler space $M_{i}$ and we use the normalisation such that the
Ricci tensor of the $i$-th manifold is $R_{ab}=\delta _{i}g_{ab}$. Then the
general solution to Einstein's field equations with cosmological constant $%
\lambda =\pm (d-1)/l^{2}$ is given by: 
\begin{equation}
F(r)=\frac{r}{\prod\limits_{i=1}^{p}(r^{2}+N_{i}^{2})^{q_{i}}}\bigg[%
\int\limits^{r}\left( \delta _{1}\mp \frac{d-1}{l^{2}}(s^{2}+N_{1}^{2})%
\right) \frac{\prod\limits_{i=1}^{p}(s^{2}+N_{i}^{2})^{q_{i}}}{s^{2}}ds-2m%
\bigg]  \label{GenF}
\end{equation}%
while the constraints on the values of the nut parameters $N_{i}$ and the
cosmological constant $\lambda $ can be expressed in the very simple form
for every $i,j=\overline{1,p}$: 
\begin{equation}
\lambda (N_{j}^{2}-N_{i}^{2})=\delta _{j}-\delta _{i}
\end{equation}

It is easy to see that if the Einstein-K\"ahler spaces are
two-dimensional, \textit{i.e.} $q_{i}=1$, we recover the solution from the
previous section.

The singularity analysis of these metrics proceeds as described in the previous section. The Euclidian section of these metrics is obtained by analytical continuation of the time coordinate and of the nut parameters. From the general expression of the function $F_E(r)$\footnote{This is deduced from (\ref{GenF}) by analytical continuation of all the nut charges $N_i\ra in_i$.} it is an easy matter to see that if the root $r_0=n_j$ where $n_j$ is the nut parameter associated with an Einstein-K\"ahler manifold $M_j$ of dimension $2q_j$ then:
\beqs
\frac{4\pi }{|F_{E}^{\prime }(n_j)|}&=&\frac{4\pi n_j(q_j+1)}{\delta_j}
 \eeqs
 otherwise, for generic roots $r_0$ we deduce that:
 \beqs
 \beta =\frac{4\pi }{|F_{E}^{\prime }(r_{0})|}&=&\frac{4\pi r_0}{\delta_1-\lambda(r_0^2-n_1^2)}
 \eeqs
These formulae are very useful in the singularity analysis of these metrics.

It is also possible to incorporate the above constraints directly in the
metric. However, this would require the manifolds involved to be
non-canonically normalised. Take for instance the $6$-dimensional Taub-NUT
fibration constructed over $S^{2}\times S^{2}$. If we normalise the spheres
such that their Einstein constant is $\delta =1$ then the constraint equation on the parameters
takes the form $\lambda (n_{1}^{2}-n_{2}^{2})=1-1=0$ and we can have a
solution with non-vanishing cosmological constant only if the nut parameters
are equal. Suppose now that we normalise the spheres such that their
Einstein constants are $\delta _{1}$, respectively $\delta _{2}$. Then
the constraint should read $\lambda (n_{1}^{2}-n_{2}^{2})=\delta _{1}-\delta
_{2}$. One way to change the Einstein constant in a general equation of
the form Ricci$(M_{i})=\lambda _{i}g(M_{i})$ is to multiply the metric $%
g(M_{i})$ by a constant factor $1/\delta _{i}$. This yields $\lambda
_{i}\delta _{i}$ as the normalised Einstein constant for the new
rescaled metric\footnote{The Ricci tensor is invariant under an overall rescaling of the metric by a
constant.}. On the other hand, recall that for a $2q$-dimensional Einstein-K\"ahler with K\"ahler form $A_{i}$, the product $(dA_{i})^{q}$ is proportional to its volume form. When rescaling the metric by $1/\delta
_{i}$ the volume form gets rescaled by a factor $1/\delta _{i}^{q}$ -- hence
we must rescale $A_{i}$ by a factor of $1/\delta _{i}$ to obtain the K\"ahler form for the rescaled metric. For spheres we should then multiply $A_{i}$ by $1/\delta _{i}$ and the metric elements $d\Omega _{i}^{2}$ by $1/\delta _{i}$, for each $i=1,2$. The expression for $F(r)$ remains
unchanged in this process\footnote{It can be read from the general form (\ref{GenF}) for general values of $\delta $'s.}.

Applying this to the six-dimensional case, we obtain 
\begin{eqnarray}
ds^{2} &=&-F(r)\left( dt-\frac{2n_{1}}{\delta _{1}}\cos \theta _{1}d\phi
_{1}-\frac{2n_{2}}{\delta _{2}}\cos \theta _{2}d\phi _{2}\right) ^{2}+\frac{%
dr^{2}}{F(r)}  \notag \\
&&+\frac{r^{2}+n_{1}^{2}}{\delta _{1}}(d\theta _{1}^{2}+\sin ^{2}\theta
_{1}d\phi _{1}^{2})+\frac{r^{2}+n_{2}^{2}}{\delta _{2}}(d\theta
_{2}^{2}+\sin ^{2}\theta _{2}d\phi _{2}^{2})
\end{eqnarray}%
while the constraint equation takes the form 
\begin{equation*}
\lambda (n_{1}^{2}-n_{2}^{2})=\delta _{1}-\delta _{2}
\end{equation*}%
Solving this equation for $\delta _{2}$ and replacing its value in the
metric we obtain: 
\begin{eqnarray}
ds^{2} &=&-F(r)\left( dt-2n_{1}\cos \theta _{1}d\phi _{1}-\frac{2n_{2}}{%
\delta_1-\lambda (n_{1}^{2}-n_{2}^{2})}\cos \theta _{2}d\phi _{2}\right)
^{2}+\frac{dr^{2}}{F(r)}  \notag \\
&&+\frac{r^{2}+n_{1}^{2}}{\delta _{1}}(d\theta _{1}^{2}+\sin ^{2}\theta
_{1}d\phi _{1}^{2})+\frac{r^{2}+n_{2}^{2}}{\delta _{1}-\lambda
(n_{1}^{2}-n_{2}^{2})}(d\theta _{2}^{2}+\sin ^{2}\theta _{2}d\phi _{2}^{2})
\end{eqnarray}%
which is a solution of Einstein field equations with cosmological constant $\lambda $ for every value of the nut parameters $n_{1}$ and $n_{2}$. Notice that now the constraint equation is already encoded in the metric, and we can for convenience scale $\delta _{1}=1$. When $n_{2}=0$ it reduces to the
cosmological $6$-dimensional Taub-NUT solution obtained previously in \cite{csrm,Page1}.

It is interesting to note that the above form of the metric allows non-singular NUTs of
intermediate dimensionality constructed over the base $S^{2}\times S^{2}$.
To see this let us notice that the absence of Misner string singularities
can be accomplished if $n_{1}/\delta _{1}$ and $n_{2}/\delta _{2}$ are 
rationally related. Specifically, we can choose for example $n_{1}=2n_{2}$ and $%
\delta _{1}=1$ while $\delta _{2}=1/2$. To satisfy these relations it is
enough to take $\lambda n_{2}^{2}=-1/6$, where $\lambda =-5/l^{2}
$ in $6$-dimensions. Then regularity of the $1$-form $(d\tau -2n_{1}/\delta
_{1}A_{1}-2n_{2}/\delta _{2}A_{2})$ requires the periodicity of $\tau $ to
be given by $\frac{8\pi n_{1}}{k\delta _{1}}=\frac{8\pi n_{2}}{k\delta _{2}}$
where $k$ is some integer. It is easy to see that we can match this
periodicity with $\frac{4\pi }{|F_{E}^{\prime }(n_{1})|}=\frac{8\pi n_{1}}{%
\delta _{1}}$ if we take $k=1$. Then there exists a nut at $r=n_{1}$ which
is completely regular -- it can be easily checked that there are no
curvature singularities in this case! We conclude that the NUT
solution of intermediate dimensionality constructed over the base space $%
S^{2}\times S^{2}$ is regular.

\section{Warped-type Fibrations and Generalized Eguchi-Hanson Solitons}

We can find a very general class of solutions of Einstein's field equations
if we use the generalised ansatz \cite{csrm,Page1} 
\begin{equation}
ds_{d}^{2}=-F(r)(dt+\sum\limits_{i=1}^{p}2N_{i}A_{i})^{2}+F^{-1}(r)dr^{2}+%
\sum\limits_{i=1}^{p}(r^{2}+N_{i}^{2})g_{M_{i}}+\alpha r^{2}g_{Y}
\end{equation}%
As before $J_{i}=dA_{i}$ is the K\"ahler form for the $i$%
-th Einstein-Kahler space $M_{i}$, $Y$ is a $q$-dimensional Einstein space
with metric $g_{Y}$ and we use the normalisation such that the Ricci tensor
of the $i$-th manifold is $R_{ab}=\delta _{i}g_{ab}$ and $R_{ab}^{Y}=\delta
_{Y}g_{ab}^{Y}$.

Then the general solution of Einstein's field equations is given by 
\begin{eqnarray}
\alpha &=&\frac{\delta _{Y}}{\delta _{1}-\lambda N_{1}^{2}}  \notag \\
F(r) &=&\frac{r^{1-q}}{\prod\limits_{i=1}^{p}(r^{2}+N_{i}^{2})^{q_{i}}}\bigg[%
\int\limits^{r}\left( \delta _{1}\mp \frac{d-1}{l^{2}}(s^{2}+N_{1}^{2})%
\right) s^{q-2}\prod\limits_{i=1}^{p}(s^{2}+N_{i}^{2})^{q_{i}}ds-2m\bigg]
\end{eqnarray}%
where the constraints on the values of cosmological constant $\lambda =\mp 
\frac{d-1}{l^{2}}$ and the nut parameters $N_{i}$ can be expressed in the
following simple form: 
\begin{equation}
\lambda (N_{j}^{2}-N_{i}^{2})=\delta _{j}-\delta _{i}
\end{equation}
for every $i,j$. For $p=1$ we recover the general solution found by L\"u, Page and Pope in ref.  \cite{Page1}.

We can treat the case $\delta _{Y}=0$ (or $q=1$) if we take the limit in
which $\lambda N_{1}^{2}=\delta _{1}$ in order to keep $\alpha $ finite in
the above expressions. In general it is not necessary to have all the nut
parameters identical, though nut parameters $N_j$ corresponding to Einstein-K\"ahler spaces that have the same Einstein constants $\delta_j$ have to be equal.

The case $q=1$ is particularly interesting to us, as it will provide a generalisation of Eguchi-Hanson metrics to arbitrary odd-dimensions \cite{EH}. For simplicity we shall work in the Euclidian sector. In this case the metric can be written as 
\begin{equation}
ds_{d}^{2}=F(r)(d\chi
+\sum\limits_{i=1}^{p}2n_{i}A_{i})^{2}+F^{-1}(r)dr^{2}+\sum%
\limits_{i=1}^{p}(r^{2}-n_{i}^{2})g_{M_{i}}+r^{2}dy^{2}  \label{EHv1}
\end{equation}
and we use $\delta_1 +\lambda n^{2}_1=0$ such that: 
\begin{equation}
F(r)=\frac{1}{\prod\limits_{i=1}^{p}(r^{2}-n_{i}^{2})^{q_{i}}}\bigg[-\lambda
\int\limits^{r}\prod\limits_{i=1}^{p}(s^{2}-n_{i}^{2})^{q_{i}}sds-2m\bigg]
\end{equation}
while the constraints on the values of the nut parameters $n_{i}$ and the
cosmological constant $\lambda $ take the form $\lambda n_{i}^{2}=-\delta
_{i}$. A positive value for the cosmological constant can still be
accommodated if we take $\delta _{i}<0$ (for instance a product of
hyperboloids, for which $\delta _{i}=-1$). Let us take for simplicity a
negative cosmological constant $\lambda =-\frac{d-1}{l^{2}}$ and let us
suppose that all the $\delta _{i}$'s are the same, \textit{i.e.} $%
\delta_i=\delta$ (for instance we can have a product of spheres or more
generally products $CP^{a_i}$ factors, for various values of $a_i$,
normalised such that their cosmological constant is $\delta$). Assume then
that the base space contains a product of $b_i$ $CP^{a_i}$ factors. Then the
dimension of the total space is $d=\sum_i2a_ib_i+3$, $n_{i}^{2}=\frac{\delta
l^{2}}{d-1}\equiv n^{2}$ and we have: 
\begin{eqnarray}
F(r) &=&\frac{1}{(r^{2}-n^{2})^{\sum_ia_ib_i}}\bigg[-\lambda
\int\limits^{r}(s^{2}-n^{2})^{\sum_ia_ib_i}sds-2m\bigg]  \notag \\
&=&\frac{1}{(r^{2}-n^{2})^{\sum_ia_ib_i}}\bigg[\frac{(r^{2}-n^{2})^{%
\sum_ia_ib_i+1}}{l^{2}}-2m\bigg]
\end{eqnarray}

It is convenient at this time to make the change of variables such that $
\rho ^{2}=r^{2}-n^{2}$: 
\begin{eqnarray}
F(\rho ) &=&\frac{\rho ^{2}}{l^{2}}-\frac{2m}{\rho ^{d-3}}  \notag \\
&=&\frac{\rho ^{2}}{l^{2}}\bigg[1-\frac{2ml^{2}}{\rho ^{d-1}}\bigg]\equiv%
\frac{ \rho ^{2}}{l^{2}}g(\rho )
\end{eqnarray}

It is now easy to see that the metric (\ref{EHv1}) with $p=\sum_ib_i$ can be
written in the following form: 
\begin{equation*}
ds_{d}^{2}=\frac{4\delta\rho ^{2}}{d-1}g(\rho )(d\chi
+\sum\limits_{i=1}^{p}A_{i})^{2}+\frac{(d-1)d\rho ^{2}}{\left(\delta+\frac{
(d-1)\rho ^{2}}{l^{2}}\right) g(\rho )}+\sum\limits_{i=1}^{p}\rho
^{2}g_{M_{i}}+\frac{l^{2}}{d-1}\left(\delta+ \frac{(d-1)\rho ^{2}}{l^{2}}%
\right) dy^{2}
\end{equation*}

Making now the change of variables $(d-1)\rho ^{2}\rightarrow \rho ^{2}$,
defining $a^{d-1}\equiv 2ml^{2}(d-1)^{\frac{d-1}{2}}$ and rescaling $y$ to
absorb the constant factor $\frac{l^{2}}{d-1}$ we eventually obtain 
\begin{eqnarray}
ds_{d}^{2} &=&\frac{4\delta \rho ^{2}}{(d-1)^{2}}\left( 1-\frac{a^{d-1}}{%
\rho ^{d-1}}\right) (d\chi +\sum\limits_{i=1}^{p}A_{i})^{2}+\frac{d\rho ^{2}%
}{\left( \frac{\rho ^{2}}{l^{2}}+\delta \right) \left( 1-\frac{a^{d-1}}{\rho
^{d-1}}\right) }+\sum\limits_{i=1}^{p}\frac{\rho ^{2}}{d-1}g_{M_{i}}  \notag
\\
&&+\left( \frac{\rho ^{2}}{l^{2}}+\delta \right) dy^{2}  \label{EHfinal}
\end{eqnarray}%
which is the most general (Euclidian) form of the odd-dimensional
Eguchi-Hanson solitons \cite{EH}, whose base space contains $b_{i}$ factors $%
CP^{a_{i}}$. The general solution whose base space contains a number of unit
curvature spheres $CP^{1}=S^{2}$ has been analysed in \cite{EH,EH1}. More
generally we can replace the $CP^{a}$ factors by arbitrary Einstein-K\"ahler manifolds $M_{i}$ normalised such that their Einstein constants are equal $\delta _{i}=\delta $ for all $i=1..p$. The parameter $\delta$ is not essential and it can be absorbed by an appropriate rescaling of the radial coordinate and redefinition of the parameter $a$. Without losing generality we can then set $\delta=1$.

It is interesting to note that while the Eguchi-Hanson solitons
constructed over Einstein-K\"ahler spaces are in general nonsingular
there are also Lorentzian section in odd-dimensions for which the curvature
singularities at the origin can be easily avoided. Take for instance the
five-dimensional metric:\footnote{This metric can be formally obtained from (\ref{EHfinal}) by setting $p=1$ and $\delta=-1$ and replacing $CP^1$ by $H^2$ (see also \cite{NB}).} 
\begin{equation*}
ds_{5}^{2}=-\frac{\rho ^{2}}{4}\left( 1-\frac{a^{4}}{\rho ^{4}}\right)
(dt-\cosh \theta d\phi )^{2}+\frac{d\rho ^{2}}{\left( \frac{\rho ^{2}}{l^{2}}%
-1\right) \left( 1-\frac{a^{4}}{\rho ^{4}}\right) }+\frac{\rho ^{2}}{4}%
(d\theta ^{2}+\sinh ^{2}\theta d\phi ^{2})+\left( \frac{\rho ^{2}}{l^{2}}%
-1\right) dy^{2}
\end{equation*}%
which is a solution of vacuum Einstein field equations with negative
cosmological constant $\lambda =-\frac{4}{l^{2}}$. In order to keep the
signature of the metric Lorentzian we must restrict the values of the radial
coordinate such that $\rho >l$. Depending on the sign of the parameter $%
a^{4} $ we can have a horizon located at $\rho =a$ and in both situations
the curvature singularity located at origin is avoided. In the limit in
which the cosmological constant vanishes, \textit{i.e.} $l\rightarrow \infty 
$, the metric describes the product of a four-dimensional Eguchi-Hanson-like
metric with a flat direction and there is no way to avoid the curvature
singularity at $r=0$ while keeping the signature of the metric Lorentzian.\footnote{Or at least allow it to be Riemannian.}

\section{Conclusions}

We have considered here higher dimensional solutions of the vacuum Einstein
field equations with and without cosmological constant. These solutions are
constructed as radial extensions of circle fibrations over even dimensional
spaces that can be factored in general as products of Einstein-K\"ahler spaces.
The novelty of our solutions is that by associating a NUT charge $N$ with
every such factor of the base space we have obtained
higher dimensional generalizations of Taub-NUT spaces that can have quite
generally multiple NUT parameters. In our work we have given the Lorentzian
form of the solutions however, in order to understand the singularity
structure of these spaces we have concentrated mainly on the their Euclidian
sections. In most of the cases the Euclidean section is simply obtained
using the analytic continuations $t\rightarrow it$ and $N_{j}\rightarrow
in_{j}$. When continuing back the solutions to Lorentzian signature the
roots of the function $F(r)$ will give the location of the chronology
horizons since across these horizons $F(r)$ will change the sign and the
coordinate $r$ changes from spacelike to timelike and vice-versa.

To render such metrics regular one follows a procedure \cite{Page} in which
the basic idea is to turn all the singularities appearing in the metric into
removable coordinate singularities. For generic values of the parameters the
metrics are singular -- it is only for careful choices of the parameters
that they become regular. In order to globally define the $1$-form $d\tau
+\sum\limits_{i=1}^{p}2n_{i}A_{i}$ we use various coordinate patches to
cover the manifold, defining the $1$-form on each patch. This can be done
consistently only if we identify $\tau$ periodically, while the nut
parameters $n_{i}$ must be rationally related. On the other hand, $r=\pm
n_{i}$ correspond to curvature singularities, unless we also require that $%
F_{E}=0$ there as well. Now by removing the possible conical singularities
at the roots of $F_{E}(r)$ (be they at $r=\pm n_{i}$ or elsewhere) we get
another periodicity for the Euclidian time $\tau $. By matching the two
periodicities we obtained for $\tau $ we get another restriction on the
value of the parameters appearing in the metric. As an example of this
analysis we have considered the $6$-dimensional Taub-NUT spaces constructed
over both $S^{2}\times S^{2}$ and $CP^{2}$. While the fibration over $CP^{2}$
is in general non-singular, we found that only the bolt solution was
non-singular for the fibration over $S^{2}\times S^{2}$, with distinct nut
parameters.

While one could think that more generally there are no regular fibrations
with distinct nut parameters over base spaces that are products of identical
factors, it turns out that this is not the case for fibrations over products
of distinct manifolds. Take for instance the $8$-dimensional metric
constructed as a fibration over $CP^{2}\times S^{2}$. If the cosmological
constant is zero we can have in general two distinct nut parameters. There
then exists a NUT of intermediate dimensionality: assuming that the nut
parameter corresponding to the $CP^{2}$ factor is $n_{1}$, while the one
corresponding to $S^{2}$ is $n_{2}$, then the periodicity of the Euclidian
time can be set to $8\pi n_{2}=12\pi n_{1}$. There exists a regular $4$-dimensional nut
located at $r=n_{2}=\frac{3}{2}n_{1}$.

As discussed in Section $3$, there exists another way to obtain NUTs of
intermediate dimensionality constructed over spaces of the same nature.
However the price to pay is the use of non-canonically normalised Einstein-K\"ahler manifolds.

We would also like to take the opportunity and comment at this point
on the existence of Misner string singularities in cases where the base
space contains $2$-dimensional hyperboloids $H^{2}$
(respectively planar geometries $T^{2}$). In the literature it is
often stated that in these cases there are no hyperbolic (respectively
planar) Misner strings \cite{Chamblin,micky1,micky2}. From the general
discussion in Section $2$ we can see that this statement is true
only if the Einstein-K\"ahler geometries $H^{2}$
(respectively $T^{2}$) are not compact. Otherwise, we find that the
integral of the $2$-form $2ndA$over closed $2$-cycles in $H^{2}$(respectively $T^{2}$) can have a finite value. This implies that the Euclidian time $\tau $ must have
(under appropriate normalization of the compact space) periodicity $8\pi n/k$, for an integer $k$; we can therefore speak about hyperbolic (planar) Misner strings.

Our construction applies more generally, yielding multiple nut-charged
generalizations of inhomogeneous Einstein metrics on complex line bundles %
\cite{csrm,Page1}. In this case we replace the Einstein-K\"ahler
manifold $M$ by a product of Einstein-K\"ahler manifolds $M_i$ with
arbitrary even-dimensions and to each such factor we associate a nut
parameter $N_i$. As is was conjectured in \cite{csrm}, we find that, quite
generally, in higher dimensions there are various constraints to be imposed
on the possible values of the cosmological constant $\lambda$, the nut
parameters $N_i$ and the values of the various $\delta$'s. These solutions
represent the multiple nut parameter extension of the inhomogeneous Einstein
metrics on complex line-bundles described in \cite{Page}. It is also
possible to cast these solutions into a different form, by explicitly
encoding the constraint conditions into the metric. However this requires us
to resort to non-canonically normalised Einstein-K\"ahler manifolds.

In Section $4$ we presented the multiple nut parameter extension of the
metrics constructed by L\"u, Page and Pope in \cite{Page1}. In this case we
replaced the Einstein-K\"ahler manifold $M$ by a product of Einstein-K%
\"ahler manifolds $M_i$ with arbitrary even-dimensions and to each
such factor we associated a nut parameter $N_i$. The case in which $Y$ is
one-dimensional is particularly interesting to us since it provided us with
the most general form of the odd-dimensional Eguchi-Hanson-type instantons
found recently by Clarkson and Mann \cite{EH}.

Leaving a more detailed study of these solutions for future work, it is
worth mentioning that our solutions can be used as test-grounds for the $
AdS/CFT$ correspondence and more generally in context of gauge/gravity
dualities. For the present solutions the boundary is generically a circle
fibration over base spaces that, being products of general Einstein-K\"ahler manifolds, can have exotic topologies. In particular, one should be able to understand the thermodynamic phase structure of such
dual field theories by working out the corresponding phase structure for our gravity solutions in the bulk.

\vspace{10pt}

{\Large Acknowledgements}

This work was supported by the Natural Sciences and Engineering Council of

Canada.

\end{document}